# Optimizing Gas Sensor Performance Using Bilayer Surface Plasmon System

Albania Urriola[1], Paolo Leonelli[1], Héctor Miranda[1], Alfredo Campos[1,2*]

[1]Laboratorio Pierre y Marie Curie, Universidad Tecnológica de Panamá, Via Centenario, Panama City, 0801, Panama, Panama.
[2]Sistema Nacional de Investigación (SNI), SENACYT, Clayton ciudad del Saber, Panama City, 0801, Panama, Panama.

*Corresponding author(s). E-mail(s): alfredo.campos@utp.ac.pa, https://orcid.org/0000-0001-7921-8309;
Contributing authors: albania.urriola@utp.ac.pa; paolo.leonelli@utp.ac.pa; hector.miranda@utp.ac.pa, https://orcid.org/0000-0001-9127-4944;

**Abstract**

The sensitivity of a gold-silver bilayer system for gas detection was investigated using angle-fixed reflectance measurements and numerical calculations based on the transfer-matrix method. Two configurations of bilayer systems were analyzed: gold on silver and silver on gold. According to calculations, the best sensitivity values were achieved with a total thickness of approximately 55 nm for the gold on silver system and approximately 65 nm for the silver on gold system. These calculations were supported by experimental results in a Kretschmann configuration for ethanol gas sensing. The bilayer systems showed a huge advantage compared to the widely used single gold thin film, however, special attention should be paid to the silver on gold configuration, as corrosion plays an important role in decreasing sensitivity.

# 1 Introduction

The search for highly efficient and fast-response sensors has led to the development of sensors based on different technologies. One of these technologies involves using light to excite phenomena in materials that are sensitive to changes in the environment. This is the case with metallic nanostructures, which exhibit the phenomenon of surface plasmon resonance (SPR) corresponding to the collective oscillation of free electrons [1, 2]. The resonance frequency of SPR depends on the surrounding medium, making it suitable for sensor applications [3]. Additionally, SPR presents strong confinement of the electromagnetic field near the nanostructure surface [4], which is also an attractive property for sensing.

Optical (bio)sensors based on SPR have been developed using metallic nanoparticles and thin films. In the latter case, the SPR is excited by the coupling of an evanescent wave with the surface plasmon wave through the use of a prism (Kretschmann [5] and Otto [6] configurations), optical fibers [7] or grating structures [8]. In metallic thin films, the SPR decays exponentially away from the metal-dielectric interface with a typical decay length of 100 to 600 nm inside the dielectric media, making it applicable for detecting large molecules. [2]. Metallic thin films have been used as sensors in multiple areas such as the pharmaceutical [9], food industry [10, 11], and environmental monitoring [12].

The use of noble metals enables thin film sensors to remain unaffected by chemical reactions with the sample tested, promoting high chemical stability. Additionally, metals like gold and silver facilitate the adhesion (functionalization) of bioactive agents for the selective detection of specific molecules [13].

Thin films made of a single metal have been widely used in the literature for sensor applications [14], however, recent studies have shown that multilayer systems can considerably improve the sensor response of the system [15–19]. These studies have focused on the use of gold or silver and a dielectric or semiconductor material. Furthermore, silver and gold metals have been coated with 2D materials such as graphene [20], black phosphorus (BP) [21] and Molybdenum disulfide ($MoS_2$) [22] revealing high performance [14]. Bimetallic systems of gold and silver have also been studied in the literature [23–25].

Prism couplers are the most used form of SPR sensor, specially the Kretschmann setup which can be found in two main configurations: spectral interrogation and angular interrogation [26]. In the former case, a white light source excites the system and the reflected signal is monitored by a spectrometer. In the second case, a monochromatic light source is used, and the reflectance is monitored while the incident angle is changing. In both cases, the reflectance curves show a minimum related to the SPR excitation. The sensitivity is then defined as the ratio between the change in resonance wavelength (or resonance angle) and the change in refractive index. The sensitivity is not the only parameter that needs to be studied to understand the performance of the SPR system. Other parameters have to be taken into account, such as the quality factor, signal-to-noise ratio, detection limit, and resolution [15, 23]. When using angular interrogation configuration, it is convenient to keep the incident angle fixed (close to resonance) and observe the change in reflectance instead of monitoring the change in resonance angle [27]. This approach is known as angle-fixed SPR reflectance [26] or intensity detection and is cost-effective, avoids the challenges of mechanical angle scanning and enables real-time measurement of reflectance. Other strategies can be implemented to avoid mechanical angle scanning, such as using a CCD camera with a collimated monochromatic beam. However, the detection is limited to the dynamic range of the CCD [26].

Another type of detection involves the phase change of the reflected wave, which is reported to be more sensitive than the change in the intensity of the reflected wave, however, it requires a more sophisticated experimental setup [28].

From the literature, the thickness optimization of bimetallic Au-Ag films has been studied in spectral [23] and angular interrogation [24, 25]. In our work, we propose plasmonic bilayer systems for gas sensing using gold and silver. We conduct a detailed study of thickness optimization to improve sensitivity in angle-fixed reflectance configuration.

## 2 Methods

### 2.1 Numerical Calculations

Numerical calculations were employed to compute the reflectance vs angle curves of bilayer systems of gold on silver (Au/Ag) and silver on gold (Ag/Au). The wavelength of light was fixed to 633 nm in all calculations. Different thickness values of gold and silver were employed ranging from 10 to 60 nm in the first layer and from 0 to 60 nm in the second layer. To carry out the calculations we used the Python programming language together with the PyLlama library [29], based on the transfer matrix approach [30]. For each curve we extracted the minimum reflectance ($R_{min}$), the resonance angle ($\theta_{min}$) and the full-width at half-maximum ($FWHM$). The sensitivity was calculated by fixing the resonance angle in air and monitoring the change in reflectance with respect to the change in refractive index. Then, the maximum of this curve was chosen as the sensitivity value which is mathematically defined as $S_R = \max\{dR/dn|_{\theta_{min\ air}}\}$, where $n$ denotes the refractive index of the medium above the bilayer system. The refractive index of the medium was changed from 1.000 to 1.020 to simulate the injection of gas. The permittivity of gold and silver was extracted from Johnson and Christy [31].

### 2.2 Sample Preparation and experimental Measurements

Bilayer films of gold on silver (Au/Ag) and silver on gold (Ag/Au) were prepared by thermal evaporation in a vacuum evaporator system (KEY Model KV-301, EE.UU.) at $10^{-6}$ torr on glass substrates. Gold and silver pellets with 99.99 % purity (RD Mathis, EE.UU.) were used. The thickness of layers was calculated from the optical transmission curve fitting procedure by using RefFit software [32], which has probed high accuracy on thickness estimation in the literature [33, 34] and has been employed in metal thin films [35]. The transmission curves were fitted with a Drude-Lorentz model with initial parameters taken from Rakic et al. [36] and Rioux et al. [37]. The curve fitting procedure is presented in Fig. S.1 and Fig. S.2 of the supporting information. The gas sensor measurements were performed in a Kretschmann configuration [5] homemade system (see the sketch of Fig. S.3 in the supporting information). A polarised laser of 632.8 nm (Thorlabs HNL050LB, EE.UU.) and a high-speed photodetector (Thorlabs DET025AL/M), EE.UU.) were used. First, the resonance angle of the bilayer system was determined without the injection of ethanol gas. Then, this angle is fixed and the reflectance is recorded over time. During the injection of ethanol gas, the reflectance vs time curve shows an increment of reflectivity during the adsorption of molecules and a decrease during desorption.

## 3 Results and Discussion

Fig. 1a presents the reflectance vs angle curves in the case of a bilayer system of gold (20 nm) on silver (30 nm) for different refractive indexes of the sensing medium from $n = 1.000$ (air) to $n = 1.010$. It is observed that, at a fixed-angle (resonance angle in air), the change in reflectance ($\Delta R$) increases significantly, which is ideal for a sensor. By monitoring the reflectance vs refractive index at a fixed-angle (black curve in Fig. 1b), a non-linear behavior was observed, with the reflectance approaching an asymptotic limit as $n$ was increased. The derivative of this curve (red curve in Fig. 1b) shows a maximum which is taken as the sensitivity value with units of /RIU, where RIU is the refractive index unit. The non-linearity of the reflectance vs refractive index curve is a disadvantage concerning the linear behavior of the resonance angle vs refractive index curve (See Fig. S.4 and S.5). However, this drawback is offset by a significant change in reflectance, making experimental measurements easier.

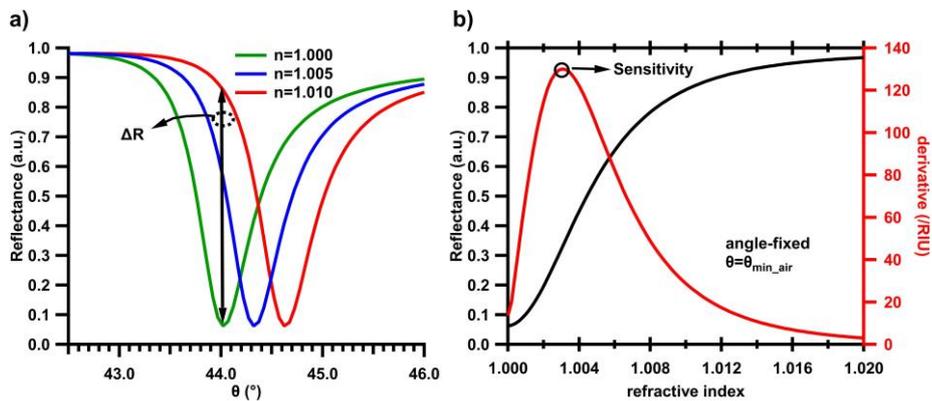

**Fig. 1 a** Reflectance vs angle curves at different refractive indexes of the surrounding medium for a gold (20 nm) on silver (30 nm) bilayer system. **b** Reflectance vs refractive index curve at a fixed-angle and its derivative, where the maximum of the derivative is defined as the sensitivity.

To know the best bilayer thicknesses for optimizing detection, the sensitivity was explored for different thickness values of the first and second layer. Fig. 2a presents the schema of a gold on silver (Au/Ag) bilayer system where the gold layer is exposed to the sensing medium. Fig. 2b presents the sensitivity map for different thickness values.

It is observed that the highest sensitivity value is obtained for a silver thickness of ~ 58 nm and a gold thickness of 0, indicating that a single silver layer shows the best sensitivity. With the addition of gold, the highest sensitivity values are obtained when the total thickness (gold thickness plus silver thickness) is approximately ~ 55 nm, and the greater the gold thickness, the lower the sensitivity. This implies that gold decreases the sensitivity, however, the incorporation of gold as second layer improves the chemical stability of the bilayer system.

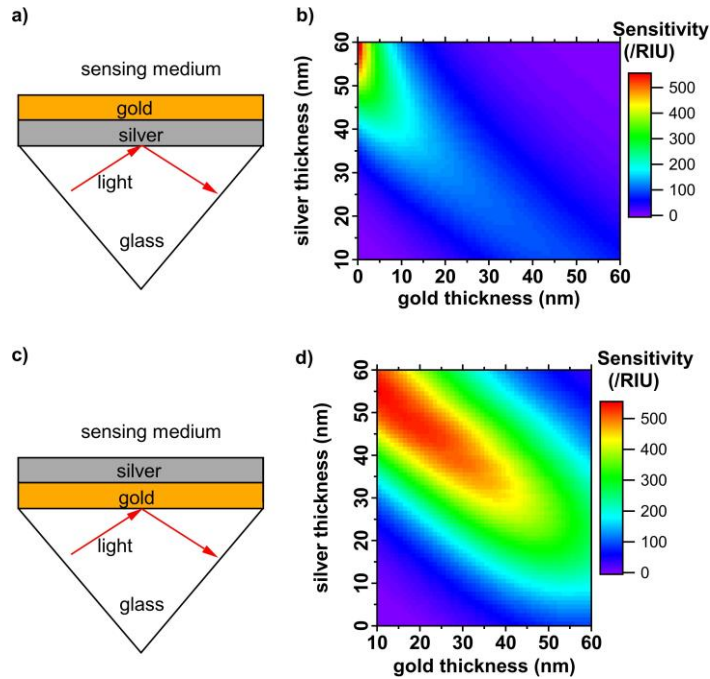

**Fig. 2 a** Sketch of a gold on silver bilayer system. **b** Sensitivity map of the gold on silver bilayer system considering a refractive index changing from 1.000 to 1.020. **c** Sketch of a silver on gold bilayer system. **d** Sensitivity map of the silver on gold bilayer system considering a refractive index changing from 1.000 to 1.020.

Fig. 2c shows a schema of a silver on gold (Ag/Au) bilayer system, where silver is exposed to the sensing medium. It is observed that the highest value of sensitivity is obtained for a gold thickness of 10 nm and a silver thickness of 55 nm. The highest sensitivity values (red zone in Fig. 2d) are achieved with different combinations of gold and silver thicknesses, particularly when the total thickness (gold thickness plus silver thickness) is around 65 nm. We can observe that the zone of high sensitivity values is larger than that of the gold on silver system presented in Fig. 2b, which is a huge advantage of the silver on gold system. Despite the fact that the silver is exposed to the sensing medium in the silver on gold system, and is therefore susceptible to corrosion, it has been proven that in core@shell systems of Au@Ag, the gold core enhances the stability properties of the silver shell [38].

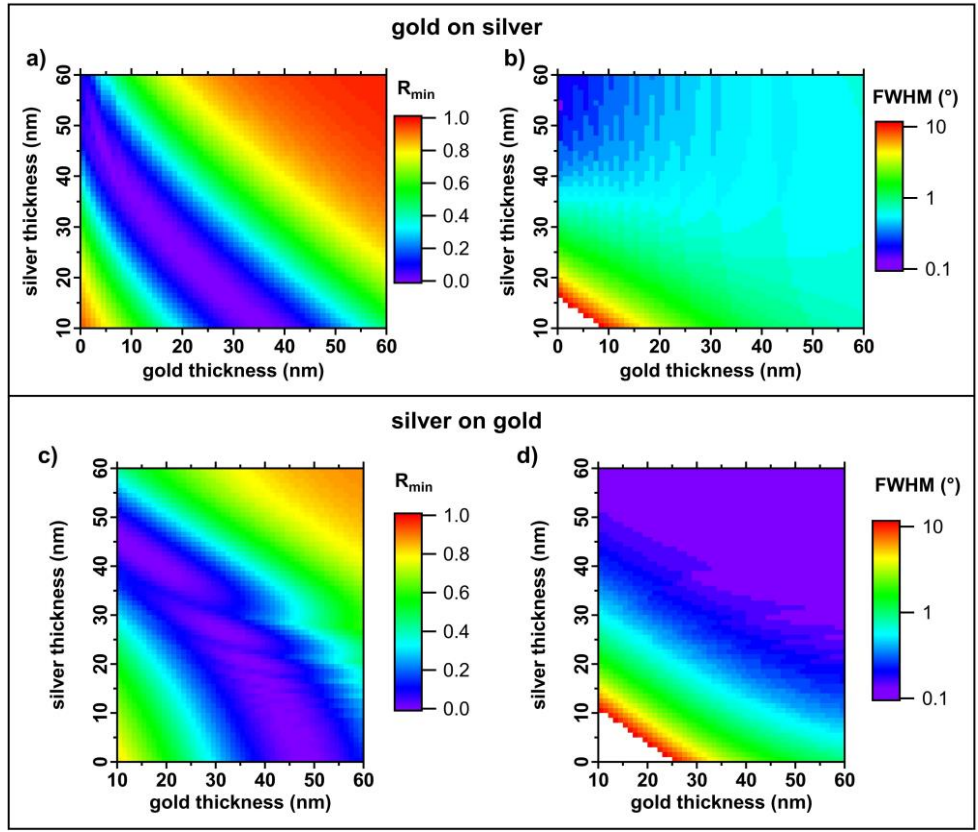

**Fig. 3** **a** Minimum of reflectance ($R_{min}$) map of gold on silver bilayer system. **b** Full-width at half-maximum ($FWHM$) map of gold on silver bilayer system. **c** Minimum of reflectance ($R_{min}$) map of silver on gold bilayer system. **d** Full-width at half-maximum ($FWHM$) map of silver on gold bilayer system

It is important to identify a relationship between the sensitivity and other important parameters such as the minimum reflectance ($R_{min}$) and the full-width at half-maximum ($FWHM$) of the reflectance vs angle curves. Fig. 3 presents the min- imum reflectance values and the full-width at half-maximum of a bilayer system of gold on silver (Fig. 3a and b) and silver on gold (Fig. 3c and d) in the presence of air as surrounding medium. We first focus on the gold on silver (Au/Ag) bilayer system. Fig. 3a shows that the lowest values of $R_{min}$ are found for different combinations of gold and silver thicknesses (purple zone). We see that the condition to find the lowest values of $R_{min}$ is obtained when the total thickness (gold thickness plus silver thickness) is approximately 50 nm, which agrees with Chen et al. [25]. Fig. 3b shows that the lowest $FWHM$ values are obtained when the silver thickness is in the range of 40-60 nm and the gold thickness is below 10 nm. The region without $FWHM$ data is related to the asymmetry of the curves and, therefore, the $FWHM$ is not possible to define. We can observe that the regions with the lowest values of $R_{min}$ and $FWHM$ presents the highest values of sensitivity (see Fig. 2b).

Now, we focus on the silver on gold (Ag/Au) bilayer system. In Fig. 3c we see that the lowest values of $R_{min}$ are found for different combinations of gold and silver thicknesses (purple zone). The condition to find the lowest values of $R_{min}$ is obtained when the total thickness (gold thickness plus silver thickness) is ~ 55 nm. Fig. 3d presents the $FWHM$ values for different combinations of gold and silver thicknesses. The lowest $FWHM$ values are obtained over a wide range of thicknesses (blue and purple zones on the graph). As discussed above, the region without $FWHM$ data is related to the asymmetry of the curves and, therefore, the $FWHM$ is not possible to define. We can observe that the region with the lowest values of $R_{min}$ and $FWHM$ presents the highest values of sensitivity (see Fig. 2d).

According to the above-presented results, The lower the value of $R_{min}$ and the narrower the $FWHM$, the greater the sensitivity.

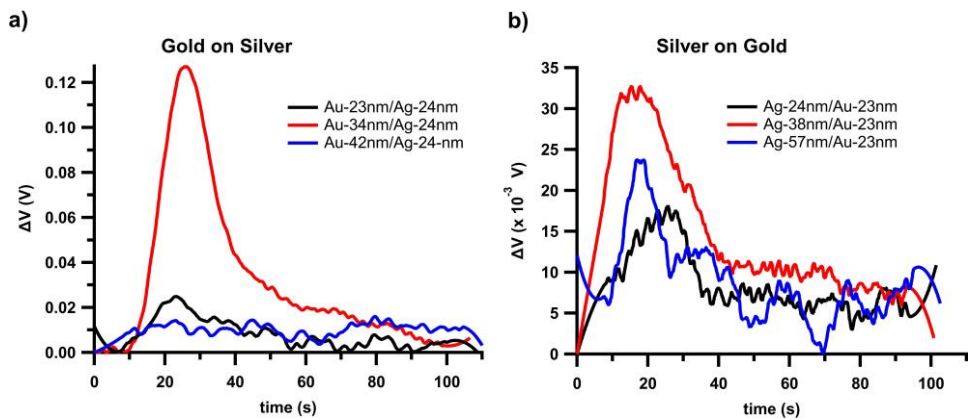

**Fig. 4** Experimental ethanol gas response over time of bilayer samples. **a** Gold on silver samples. **b** Silver on gold samples. The carrier gas flux was fixed to 2 lpm and the injected volume of ethanol was 200 $\mu$L.

The sensor response of bilayer systems were studied experimentally through the injection of ethanol gas in a Kretschmann configuration. The gold on silver samples prepared were Au-23nm/Ag-24nm, Au-34nm/Ag-24nm and Au-42nm/Ag-24nm and their sensor responses are presented in Fig. 4a, where the best response corresponds to the sample Au-34nm/Ag-24nm. On the other hand, the silver on gold samples prepared were Ag-24nm/Au-23nm, Ag-38nm/Au-23nm and Ag-57nm/Au-23nm and their sensor responses are presented in Fig. 4b, where the best response corresponds to the sample Ag-38nm/Au-23nm. To understand why some samples exhibit better sensor response than others, we compared the experimental results with the calculations in Fig. 5. Fig. 5a shows the simulations alongside the experimental samples of the gold on silver bilayer system. It is observed that the sample Au-34nm/Ag-24nm is closer to the line of the best sensitivity, which explains the better performance of this sample. On the other hand, Fig. 5b shows the simulations alongside the experimental samples of the silver on gold bilayer system. It is observed that the sample Ag-38nm/Au-23nm is closer to the line of the best sensitivity, which explains the better performance of this sample.

According to calculations, the Ag-38nm/Au-23nm sample has higher sensitivity compared to Au-34nm/Ag-24nm. However, the opposite occurs in the experiment. This could be due to the possible corrosion of silver in the silver on gold samples, resulting from exposure to the environment. The literature extensively discusses the corrosion of silver [39–41] and the use of protective layers is promoted. The effect of silver corrosion on the silver surface morphology and sensitivity is discussed in Fig. S.6 and S.7, respectively.

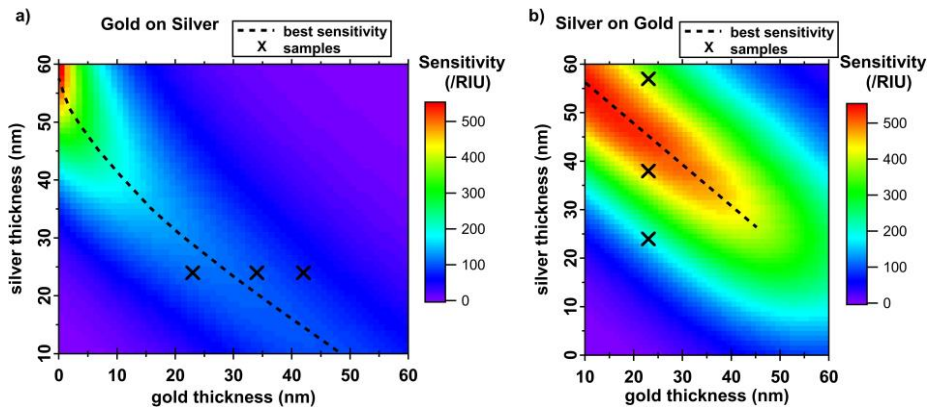

**Fig. 5 a** Sensitivity map of a gold on silver configuration alongside with experimental data. **b** Sensitivity map of a silver on gold configuration alongside with experimental data. In both **a** and **b**, the dotted lines represent the points with the highest sensitivity values, while the crosses represent the experimental samples.

## 4 Conclusions

Gas detection sensitivity of bilayer systems of gold on silver (Au/Ag) and silver on gold (Ag/Au) was studied by numerical calculations based on the transfer-matrix approach. The sensitivity was extracted from the maximum of the derivative of the reflectance vs refractive index curve at angle-fixed and it was computed for different combinations of thickness values. High sensitivity values were associated with low minimum reflectance values and sharp peaks in the reflectance versus angle curves in air. Au/Ag configuration shows a high sensitivity when the total thickness of the system was around 55 nm, while in the case of Ag/Au when the total thickness was around 65 nm. To experimentally validate these findings, bilayer samples were prepared using thermal evaporation and their sensor responses were studied in the Kretschmann configuration by injecting ethanol gas. The best sensor response performances were obtained for Au-34nm/Ag-24nm and Ag-38nm/Au-23nm samples, which agree with the calculations. This research focused on optimizing the thicknesses of bilayer systems of gold and silver to enhance gas sensor response. Special attention must be given to the Ag/Au configuration due to the risk of silver corrosion, which can degrade sensitivity.

**Supplementary information.** Details on thickness measurements by transmission spectra fitting, experimental setup for gas sensor measurements, reflectance vs refractive index behavior and the effect of corrosion on silver film morphology and sensitivity are presented.

**Acknowledgements.** Thanks to SENACYT and CEMCIT-AIP for their financial support through funding projects FID18-066 and APY-NI-2022-15. Alfredo Campos thanks the Sistema Nacional de Investigación (SNI) de Panamá for financial support. Finally, thank you to Alexander Wittel for help with the English revision.

## Declarations

- Funding
  This work is supported by the SENACYT under the project FID18-066, project APY-NI-2022-15 and the Sistema Nacional de Investigación (SNI) de Panamá.
- Competing interests
  The authors have no relevant financial or non-financial interests to disclose
- Ethics approval and consent to participate
  This study does not involve humans, animals or biological material. No ethical approval is required
- Consent for publication
  This study does not involve humans, animals or biological material. No consent for publication is required
- Data availability
  The datasets generated and analyzed in this article are available from the corresponding author on reasonable request.
- Materials availability
  Not applicable
- Code availability
  The Python codes performed in this article are available in a public github repository
  https://github.com/alfredo2711/Thickness-optimization
- Author contribution
  All authors contributed to the study, conception and design. Bilayer films were fabricated by Albania Urriola in an evaporator system optimized by Héctor Miranda. Optical transmittance measurements were carried out by Albania Urriola and the curve fitting procedure was performed by Albania Urriola and Paolo Leonelli. The